\documentclass[prb,twocolumn,preprintnumbers,superscriptaddress]{revtex4-2}
\usepackage{graphicx}
\usepackage{dcolumn}
\usepackage{bm}
\usepackage{graphicx}  
\usepackage{dcolumn}   
\usepackage{bm}        
\usepackage{verbatim}   
\usepackage[usenames]{color}  
\usepackage{stmaryrd}
\usepackage{float}
\usepackage{units}
\usepackage{textcomp}
\usepackage{textcomp}
\usepackage{amsmath}
\usepackage{commath}
\usepackage{amssymb}
\usepackage{esint}
\usepackage{ae,aecompl}
\usepackage[colorlinks=true,linkcolor=blue,citecolor=blue]{hyperref}
\usepackage{subfigure}

\begin{document}

\title{Magneto-optical properties of electron gas in a chain of planar quantum rings: \\ Effect of screening on the signature of quantum phase interference}

\author{Armen Harutyunyan}
\email{armen.harutyunyan@ysu.am}
\affiliation{Department of Solid State Physics, Yerevan State University, Alex Manoogian 1, 0025 Yerevan, Armenia}

\author{Vram Mughnetsyan}
\email{vram@ysu.am}
\affiliation{Department of Solid State Physics, Yerevan State University, Alex Manoogian 1, 0025 Yerevan, Armenia}

\author{Albert Kirakosyan}
\email{kirakosyan@ysu.am}
\affiliation{Department of Solid State Physics, Yerevan State University, Alex Manoogian 1, 0025 Yerevan, Armenia}

\author{Vidar Gudmundsson}
\email{vidar@hi.is}
\affiliation{Science Institute, University of Iceland, Dunhaga 3, IS-107 Reykjavik, Iceland}

\begin{abstract}
The equilibrium properties and interminiband transitions for Hartree-interacting two-dimensional electron gas in a one-dimensional chain of planar quantum rings subjected to a transverse homogeneous magnetic field are examined theoretically and numerically.
The proposed analytical models for the external modulation potential and the basis wave-function reflect the symmetry and the topology of the system allowing to get high accuracy results in  comparatively short computational times.
The calculated dependencies of the electronic band structure versus magnetic field show two detached minibands in the region of low energies, while higher minibands manifest multiple crossings and anticrossings.
The existence of highly degenerate energy levels (referred to as miniband nodes) for certain values of magnetic field is revealed.
The miniband nodes are preserved when taking into account the Coulomb interaction of the electrons.
The interaction leads to an up-shift and a broadening of the minibands, as well as to a shift of the nodes toward the larger values of the magnetic field.
The miniband nodes have their signature in the magnetization of the electron gas and in the interminiband transitions.
The number of electrons per a unite cell of the system has a strong impact on the current density distribution, magnetization and the oscillator strength.
The obtained results open new opportunities for a flexible manipulation of magneto-optical properties of future devices operating in far-infrared and THz regimes.

\end{abstract}
\maketitle

\section{INTRODUCTION}
The development of nanotechnology has been advancing for decades and gained acceleration in the 21st century.
Excellence in the fabrication of two dimensional (2D) hybrid materials with better characteristics is anticipated to result from intensive research of 2D heterostructures.
This enhanced evaluation might open new opportunities for the synthesis of 2D materials and the creation of devices that are more effective than traditional ones in various sectors of application \cite{KUMBHAKAR}.
Besides their technological importance, two-dimensional electron systems in semiconductor heterostructures have served as important test beds for advancing the understanding of quantum many-body methods and approaches in condensed matter theory
\cite{RevModPhys.54.437}.
Fundamental to this is the ability to change their electron density or modulate it spatially into arrays of quasi-one- or zero-dimensional electron systems.
2D electron gas (EG) in a perpendicular homogeneous magnetic field and a periodic lateral superlattice is known for its fractal energy spectrum originating from quantum phase interference effects, the Hofstadter butterfly
\cite{PhysRevB.14.2239, PGHarper_1955, azbel1964energy, PhysRev.180.633, MANSOURY2023129115}.
The screening of this spectrum has been investigated at the levels of Hartree
\cite{PhysRevB.54.R5223, PhysRevB.52.16744}
approximation and the local spin density functional theory.
In the latter case the effect of photon cavity on the magnetization and the real-time non-linear response of the system is considered as well
\cite{PhysRevB.106.115308, https://doi.org/10.1002/andp.202300274, PhysRevB.108.115306}
in the frame of a quantum electrodynamical density functional theory.
Another quantum phase effect with a plenty of possible applications is the so called Aharonov-Bohm effect
\cite{PhysRev.115.485}.
This effect arises from the fact that, in quantum mechanics, the canonical formalism is essential, whereas in classical mechanics, the fundamental equations of motion can always be expressed solely in terms of the field.
Nanoscale sized semiconductor rings, which are now commonly called Aharonov-Bohm quantum rings (QR), are among other quantum systems used for experimental studies of the renowned discovery
\cite{FominBook}.
Few-electron QRs with a radial size of $10-20$nm are now easily fabricated
\cite{FominBook}.
The mean free path of particles confined in these nanostructures exceeds the ring circumference, resulting in the self-interference effects.
The influence of the field potentials upon this interference in the regions with vanishing field magnitudes is a direct evidence of the Aharonov-Bohm effect present in QRs.
Persistent currents, the quantum Hall effect, system magnetization, and other intriguing phenomena have been extensively investigated in systems composed of QRs because of their unique geometrical and topological properties and their importance in the development of new-generation quantum devices
\cite{daCosta_2017}.

Great efforts have been devoted to achieve vertical and lateral alignment of QRs.
For example, stacking of three InGaAs/GaAs QRs is demonstrated to provide a broad-area laser
\cite{FerranSurez_2004}.
In QR complexes and stacks, novel correlations show up, which allow for a
control over their electronic and magnetic properties. One-dimensional (1D)
ordered QR-chains have been fabricated on a quantum-dot superlattice template using molecular beam epitaxy, strain field engineering and partial capping technics
\cite{Wu2012}.
It has been shown that the combination of the phase coherence effects related with angular and linear motions of an electron in a 1D chain of 2D QRs may lead to highly degenerate quantum states due to the collapse of minibands for some values of the transverse magnetic field, which can be flexibly controlled
\cite{MANSOURY2022128324}.
At the same time the number of electrons in each QR is a crucial parameter defining the period of the Aharonov-Bohm oscillations. Exact diagonalization method revealed the fractional Aharonov–Bohm effect of a few-electron system in a one-dimensional QR taking into account spin, disorder and the Coulomb interaction
\cite{Niemela_1996}.
A great challenge for the theory — to find analytical solutions for quantum states in QRs — was addressed for two electrons on a one dimensional QR for particular values of the ring radius \cite{PhysRevLett.108.083002}.
Many - electron QRs have been studied using a number of versions of Density Functional Theories
\cite{PhysRevB.79.121305, PhysRevB.62.10668}.

In the light of above mentioned, the study of magneto - optical properties of a Coulomb - interacting 2D EG in a 1D chain of QRs is of great fundamental and practical interest.
In our previous paper \cite{MANSOURY2022128324}
we have shown that a 1D chain of planar QRs exhibits unique optical properties which are directly connected with the existence of nodes (highly degenerate states) in the electron band structure.
This phenomenon being directly coupled to magnetic Berry phases
\cite{Phase},
can be notably affected by interparticle interactions.
In this paper we focus on the equilibrium properties and interminband transitions of a 2D Hartree-interacting EG in a 1D chain of planar QRs exposed to a transverse homogeneous magnetic field.
We discuss in details the origins of the miniband energy degeneration and the effect of the Coulomb screening on the miniband nodes.
Further, we provide a study on the electron density, persistent currents and magnetization of the system - an equilibrium characteristic which can be measured directly in experiment.
We have shown that miniband nodes have their obvious signature in the interminiband transitions' intensity and is largely influenced by the screening.
The paper is organized as follows: a brief description of the theoretical framework is presented in section \ref{theory}, section \ref{discussion} is devoted to the discussion of the results, in section \ref{summary} the summary of the main results is presented and finally, the acknowledgements are in section \ref{acknowledgements}.

\section{Theory}
\label{theory}
\begin{figure}
\centerline{\includegraphics[width=9cm]{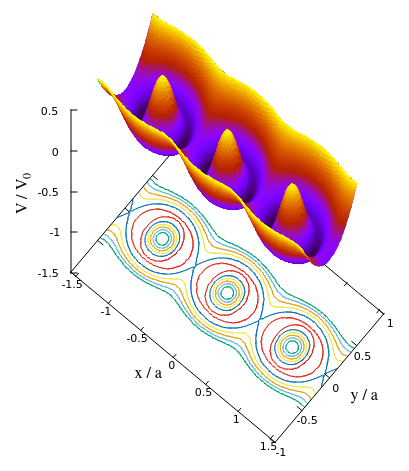}}
\caption{(Colour on-line) Modulation potential profile and the contour plot for the chain of QRs according to Eq. (\ref{Vpot}) for the following values of parameters: $v_1 = 1$, $v_2 = 1.5$, $v_3=0.05$, $\gamma_1=0.15$, $\gamma_2 = 0.45$.}
\label{chain}
\end{figure}
We consider an effective 2D motion of electrons (in $xy$ plane) in an external modulation potential which is periodic in the $x$ direction, while the electrons are confined in the $y$ direction.
A physical structure where such a modulation can be realized is a 1D SL of quasi-2D QDs or QRs,  periodically arranged in the $x$ direction and surrounded by an insulator from both sides.
We are using here an analytic model for the modulation potential which can be flexibly tuned by the parameters governing the period of the chain, the confinement in the transverse direction as well as the barriers' height in the centers and in between the QRs
\begin{equation}
\begin{aligned}
V_{\mathrm{ext}}(x, y) = V_0 & \left[ -v_1 \cdot \cos^2\left(gx/2\right) \exp{(-\gamma_1\left(gy\right)^2)} \right. \\
& \left. + v_2 \cdot \cos^4\left(gx/2\right) \exp{(-\gamma_2\left(gy\right)^2)} \right. \\
& \left. + v_3 \cdot \left(gy\right)^2 \right],
\end{aligned}
\label{Vpot}
\end{equation}
where $g=2\pi/a$ is the primitive vector of reciprocal lattice of a chain with a period $a$. In the numerical calculations we have chosen the following values for the parameters: $v_1 = 1$, $v_2 = 1.5$, $v_3=0.05$, $\gamma_1=0.15$ and $\gamma_2 = 0.45$. This choise of the parameters results in an obvious ring topology in each unit cell (UC) of the SL (see Fig.\ \ref{chain}).
The whole structure is exposed to a transverse homogeneous magnetic field $\vec{B}$=$B\hat{k}$, where $\hat{k}$ is the unit vector in the $z$ direction.

In the effective mass approximation the Hamiltonian of the system of Hartree-interacting electrons is
\begin{equation}
  H=\frac{1}{2m}\left(\vec{p}+\frac{e\vec{A}}{c}\right)^2+V_{\mathrm{ext}}(\vec{r})+V_\mathrm{H}(\vec{r}) \pm g^{*} \mu_{\mathrm{B}}^{*} B / 2,
\label{Hamiltonian}
\end{equation}
where the last term is the Zeeman energy with effective Land\'{e} factor $g^{*}$, and the Bohr magneton $\mu_{\mathrm{B}}^{*}$. \( V_{\mathrm{ext}}(\vec{r})\) is the periodic lattice potential defined by Eq.\ (\ref{Vpot}) and
\begin{equation}
V_\mathrm{H}(\vec{r}) = \frac{e^2}{\varkappa} \int_{\mathbb{R}^2} \frac{\Delta n(\vec{r'})}{|\vec{r} - \vec{r'}|} d\vec{r'}
\label{VH}
\end{equation}
is the Hartree potential, where
\begin{equation}
\Delta n(\vec{r}) = n(\vec{r}) - n_b(y),
\label{Dn}
\end{equation}
$n(\vec{r})$ is the electron density, and $n_b$ is the positive background charge density which provides the neutrality of the system:
\begin{equation}
n_b(y) = \frac{1}{a}\int_{-a/2} ^{a/2} n(\vec{r})dx.
\label{nb}
\end{equation}
In Eqs.\ (\ref{Hamiltonian})-(\ref{nb}) $\vec{p}=-i\hbar\vec{\nabla}$ is the momentum operator, $\vec{A}$ is the magnetic vector potential, $m$ is the electron effective mass, $e$ is the elementary charge, $\varkappa$ is the dielectric constant, and $c$ is speed of light.
Note, that the background charge is a function of the $y$ coordinate exponentially vanishing for large enough values of $y$.

Within the framework of the Landau gauge, $ \vec{A}(-By,0,0) $, the one-electron wave function can be expressed as
\begin{equation}
 \Psi_{j,k}(x,y)=\frac{\exp{(ikx)}}{\sqrt{a}}\sum\limits_{n,G} C_{n,G}^{(j,k)} \exp{(iGx)}f_{n}(\xi_{k,G} (y)),
\label{Psi}
\end{equation}
where
\begin{equation}
f_{n}(\xi)=\frac{\exp{\left(-\xi^{2}/2\right)}}{\sqrt[4]{\pi l^{2}_{B}}\sqrt{2^n \cdot n!}}  H_{n}\left(\xi\right),
\label{fn}
\end{equation}
\begin{equation}
\xi_{k,G}(y)=(y/l_{B}-l_{B}(k+G))
\label{xi}
\end{equation}
and $H_{n}(\xi)$ is the Hermite polynomial of the $n$-th order. $l_{B}=(c\hbar/eB)^{1/2}$ is the magnetic length, $G=gn_{x}$, with integer $n_{x}$, and $\hbar k$ is the one-dimensional quasi-momentum.
Note, that the wave function (\ref{Psi}) reflects both the periodicity of the considered system and the effects of the magnetic field.

The electron density can be expressed as
\begin{eqnarray}
n(\vec{r}) = \frac{a}{2\pi} \sum_{j,s} \int_{-\frac{\pi}{a}}^{\frac{\pi}{a}} f_\mathrm{FD}(E(j,s,k))\left|\Psi_{j,k}(\vec{r})\right|^2 dk,
\label{ne}
\end{eqnarray}
where $f_\mathrm{FD}(E)$ is the Fermi-Dirac distribution function, $E$ represents the energy spectrum of the Hartree - interacting electrons, $j$ and $s$ are the miniband and the spin quantum numbers, respectively.
The one dimensional periodicity of the problem allows one to diagonalize the Hamiltonian for each Hartree iteration in the space reciprocal to $x$. The nondiagonal matrix elements are
\begin{equation}
\begin{split}
& V(k,G,G',n,n') =  \int_{-\infty}^{\infty}
f_{n'}(\xi_{k,G'}(y))f_{n}(\xi_{k,G}(y)) \\
& \times (v_{\mathrm{ext}}(G'-G,y) + v_{\mathrm{H}}(G'-G,y)) dy,
\end{split}
\label{VkGGnn}
\end{equation}
where
\begin{widetext}
\begin{equation}
\begin{split}
v_{\mathrm{ext}}(G,y) = V_0 \Bigg\{ & -v_1 e^{-\gamma_1\left(gy\right)^2}\cdot\frac{1}{4}(2\delta_{nx,0} + \delta_{nx,1} + \delta_{nx,-1}) \\
& + v_2 e^{-\gamma_2\left(gy\right)^2}\cdot\frac{1}{4}\left(\frac{3}{2}\delta_{nx,0} + \delta_{nx,1} + \delta_{nx,-1}+\frac{\delta_{nx,2} + \delta_{nx,-2}}{4}\right) \\
& + v_3 \left(gy\right)^2\cdot\delta_{nx,0} \Bigg\},
\end{split}
\label{vGy}
\end{equation}
\end{widetext}
and
\begin{equation}
v_{\mathrm{H}}(G,y) = \frac{2e^2}{\varkappa} \int_{-\infty}^{\infty} \Delta n({G}, {y'})\cdot K_0(|G||y-y'|)d{y'}
\label{vH}
\end{equation}
are the one-dimensional Fourier transforms of the external and the Hartree potentials, respectively, $\delta_{i,j}$ is the Kronecker symbol and
$K_0(\zeta)$ is the zero order Bessel function of the second kind.
The Fourier transform of the electron density is
\begin{equation}
\begin{split}
n(G,y) &= \sum_{n,n',G',j,s}f_{n}(\xi_{k,G}(y)) f_{n'}(\xi_{k,G'-G}(y)) \\
&\quad \times C_{n',G'-G}^{(j,k)*} C_{n,G}^{(j,k)} f_{FD}(E(j,s,k)).
\end{split}
\end{equation}
We discuss here equilibrium quantities, namely the mean value of the persistent current density:
\begin{equation}
\langle \vec{j}_{n,k}(\vec{r}) \rangle = -(e/2)\langle (\hat{\vec{v}}|n,k\rangle \langle n,k|+|n,k\rangle \langle n,k|\hat{\vec{v}}) \rangle,
\end{equation}
and the mean value of the magnetization per the unit cell of the QR chain
\begin{equation}
\langle M \rangle = \frac{1}{2ca} \int_{UC}(\vec{r} \times \langle \vec{j}_{n,k}(\vec{r}) \rangle)d^{2}r.
\end{equation}
To examine the screening effect on the quantum transitions between the $1$st and the $2$nd minibands we calculate the mean value of their oscillator strength (OS) defined as follows \cite{demtröder2002laser}
\begin{equation}
\begin{split}
    O_{1,2} &= \frac{a^{2}}{(2 \pi)^{2}} \sum_{s_{1},s_{2}}
    \int_{-\frac{\pi}{a}}^{\frac{\pi}{a}} dk_{1}dk_{2}
    O_{1,2}(k_{1},k_{2},s_{1},s_{2}) \\
& \times  f_\mathrm{FD}(E(1,s_{1},k_{1}))
    (1-f_\mathrm{FD}(E(2,s_{2},k_{2})))
    ,
\end{split}
\end{equation}
where
\begin{equation}
\begin{split}
   O_{1,2}(k_{1},k_{2},s_{1},s_{2}) &=
   \frac{2m(E(2,s_{2},k_{2})-E(1,s_{1},k_{1}))}{3 \hbar^{2}} \\
&   \times |\langle 2, k_{2} |\vec{r}| 1, k_{1} \rangle|^{2}
\end{split}
\end{equation}
is the oscillator strength between an initial state in the first miniband (with values of quasimomentum $k_{1}$ and spin $s_{1}$) and a final state in the second miniband (with $k_{2}$ and $s_{2}$).

\section{Discussion}
\label{discussion}
\begin{figure}
\centerline{\includegraphics[width=0.5\textwidth]{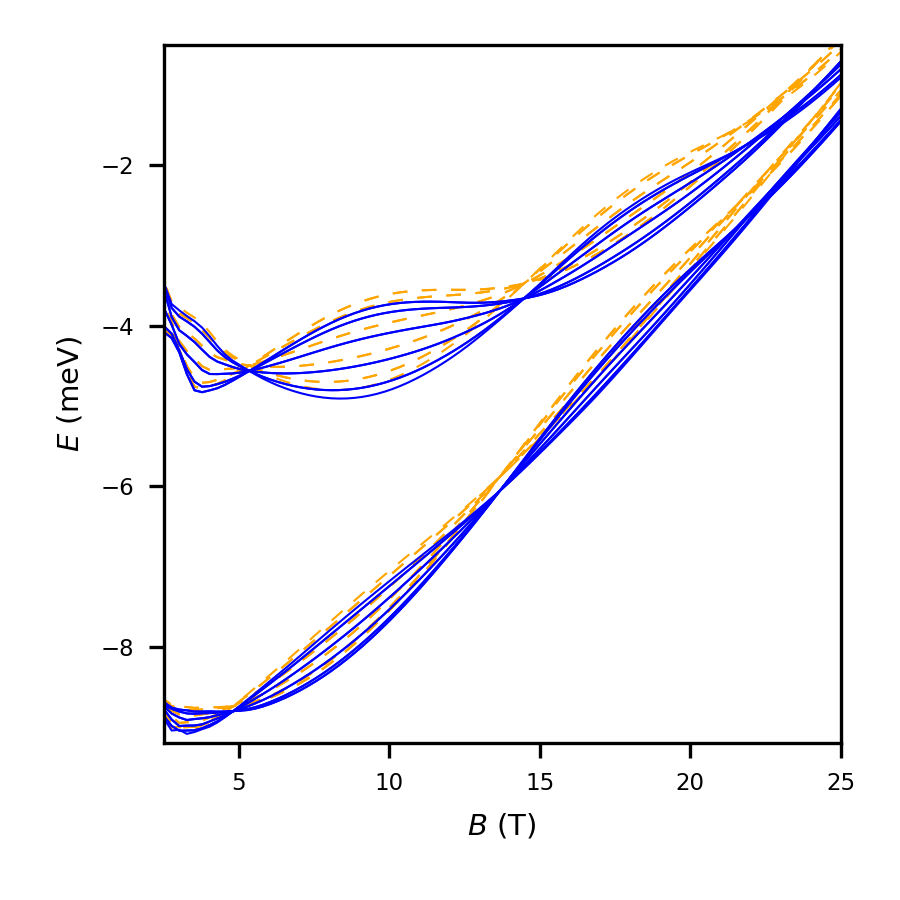}}
\caption{The lowest two miniband energies as functions of magnetic field induction for non-interacting electrons. The solid lines correspond to spin-up, wile the dashed ones to spin-down states.}
\label{Enonint1}
\end{figure}
\begin{figure}
\centerline{\includegraphics[width=0.4\textwidth]{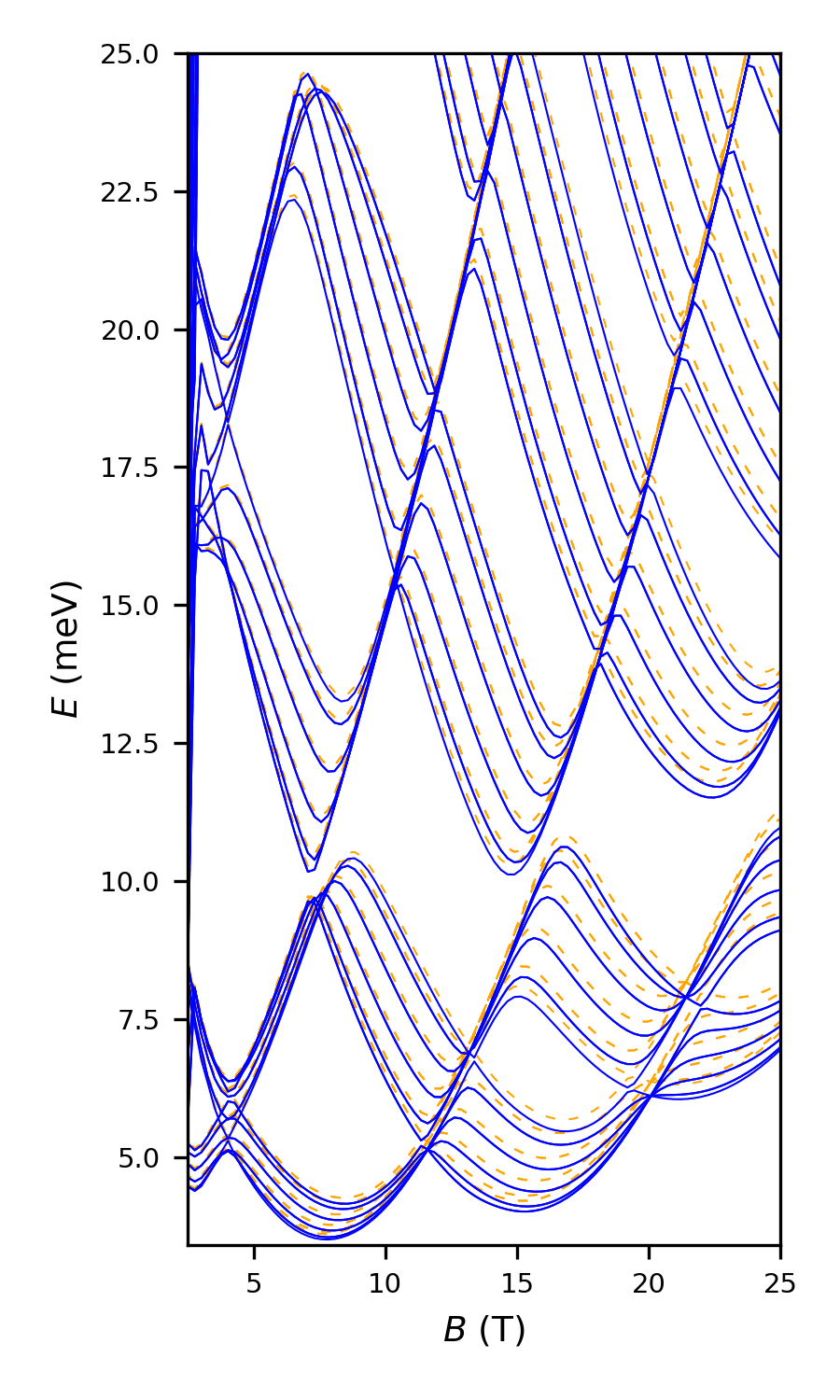}}
\caption{The 4-th to 6th miniband energies as functions of magnetic field for non-interacting electrons. The solid lines correspond to spin-up, wile the dashed ones to spin-down states.}
\label{Enonint2}
\end{figure}
\begin{figure*}
    \centering
    \begin{subfigure}
        \centering
        \includegraphics[width=0.45\textwidth]{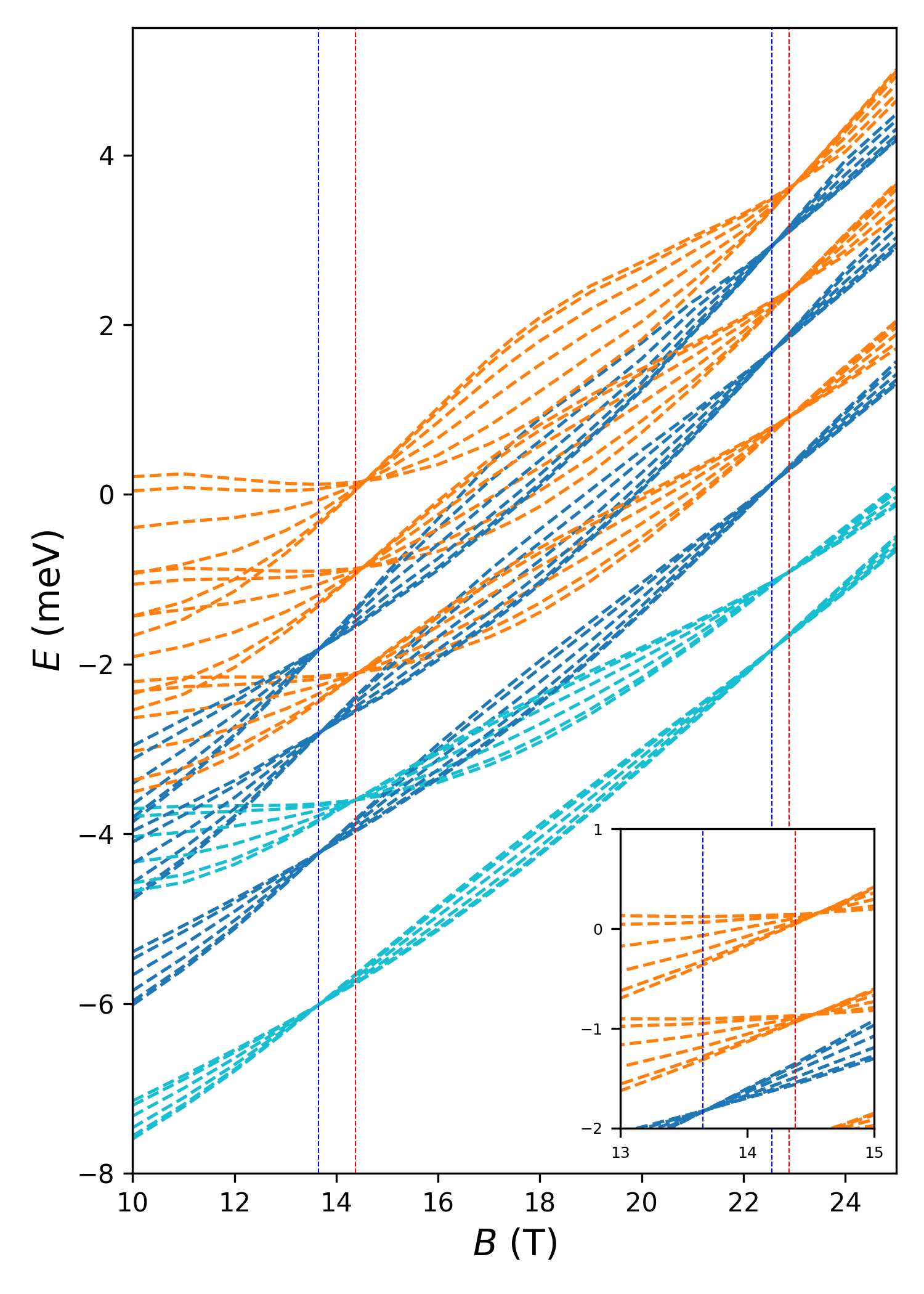}
    \end{subfigure}
    \quad 
    \begin{subfigure}
        \centering
        \includegraphics[width=0.45\textwidth]{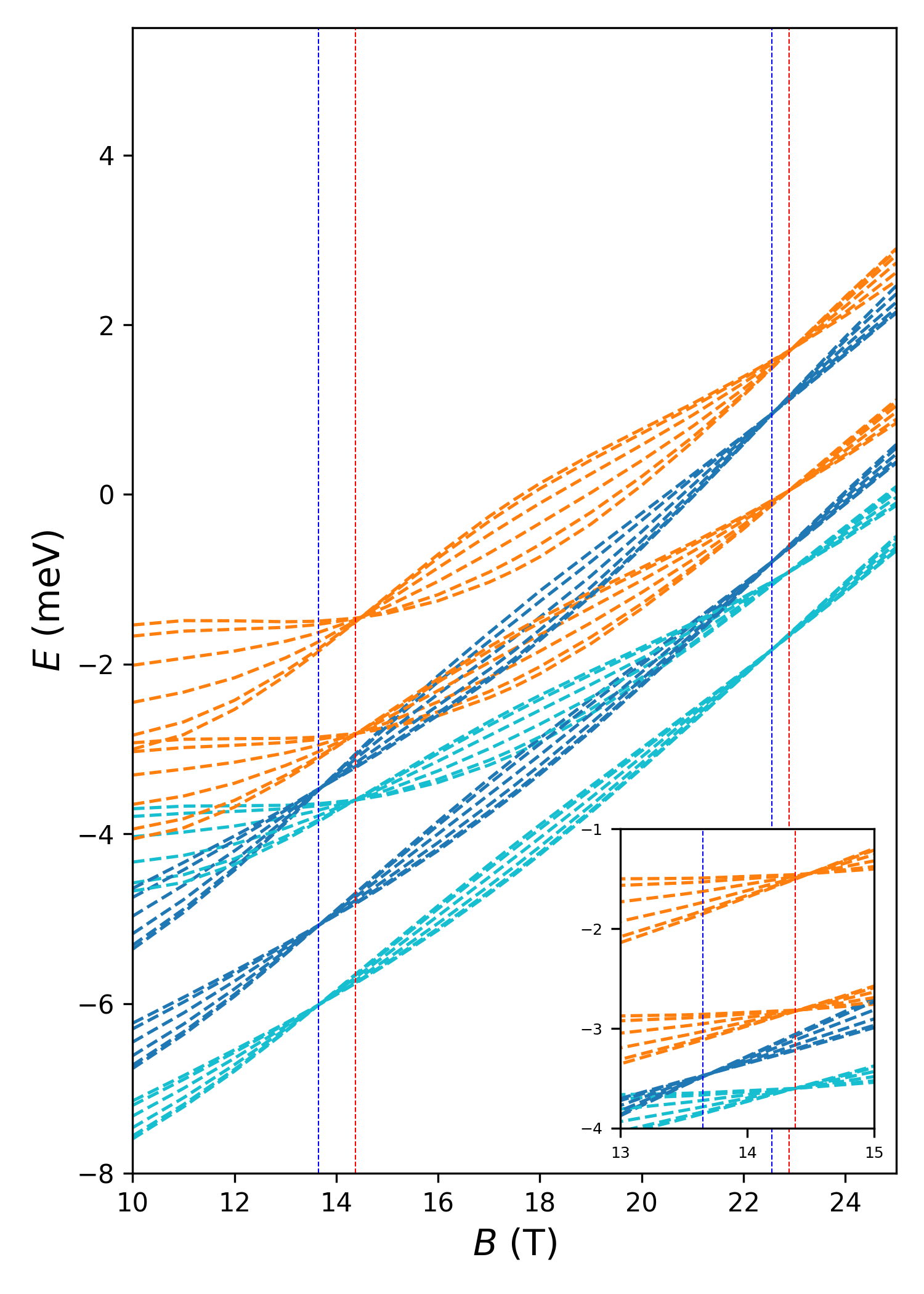}
    \end{subfigure}
    \caption{The lowest two miniband energies for spin-up states as functions on magnetic field for integer (left panel) and half an integer (right panel) numbers of electrons per unit cell. The dark-blue dotted lines correspond to noninteracting electrons.}
    \label{Eint}
\end{figure*}
\begin{figure*}
    \centering
    \begin{subfigure}
        \centering
        \includegraphics[trim=10 40 25 40, clip, width=0.31\textwidth]{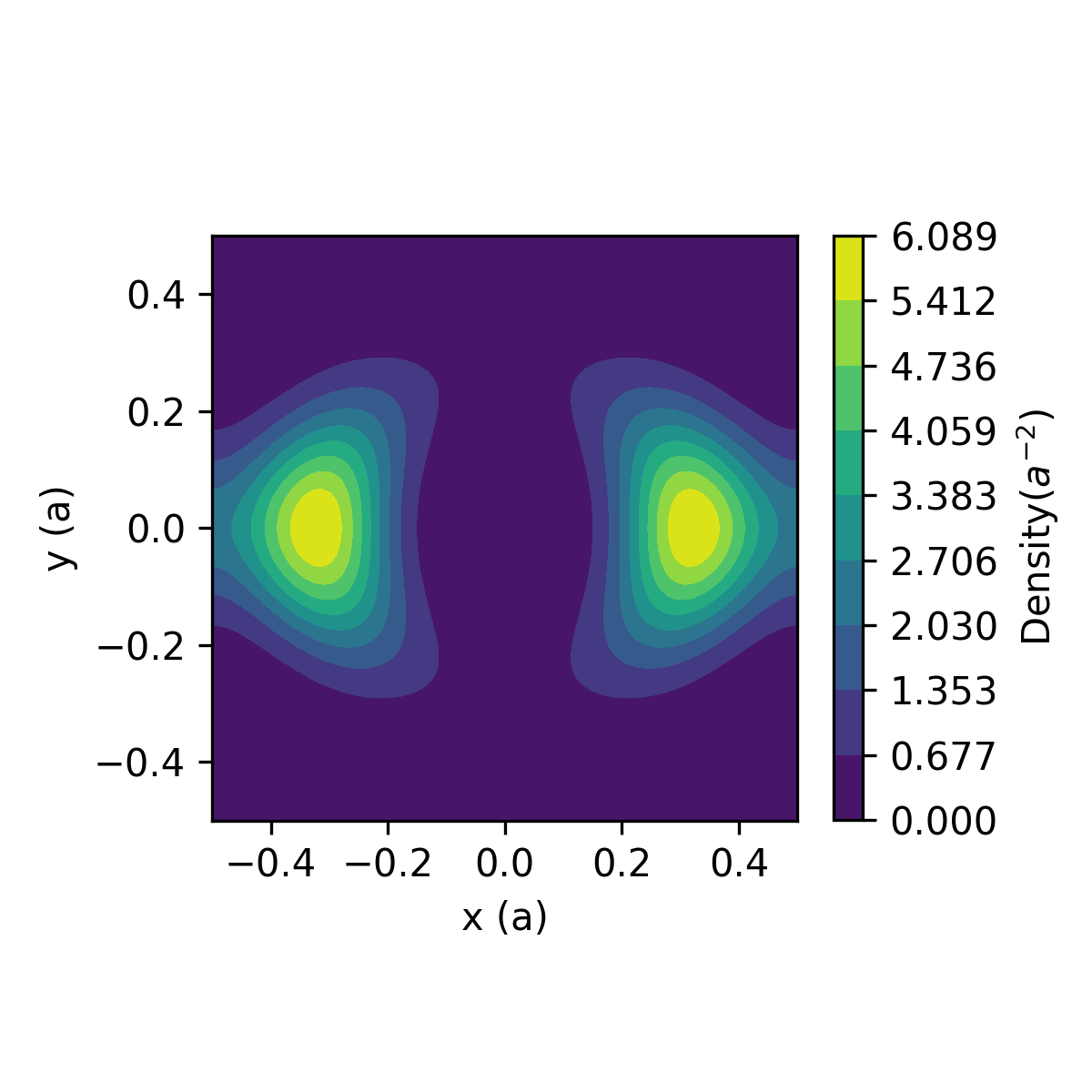}
    \end{subfigure}
    \quad 
    \begin{subfigure}
        \centering
        \includegraphics[trim=10 40 25 40, clip, width=0.31\textwidth]{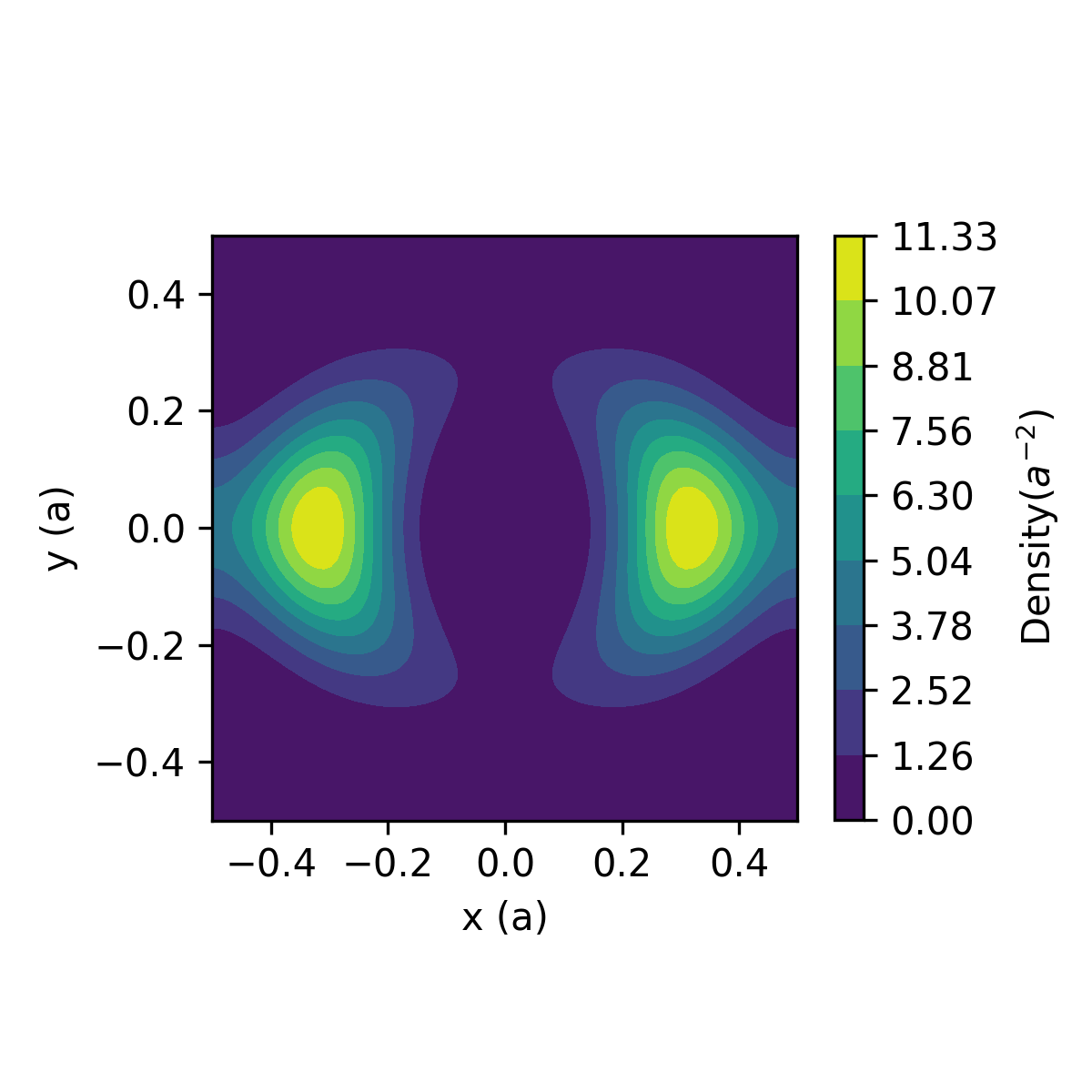}
    \end{subfigure}
    \quad
    \begin{subfigure}
        \centering
        \includegraphics[trim=10 40 25 40, clip, width=0.31\textwidth]{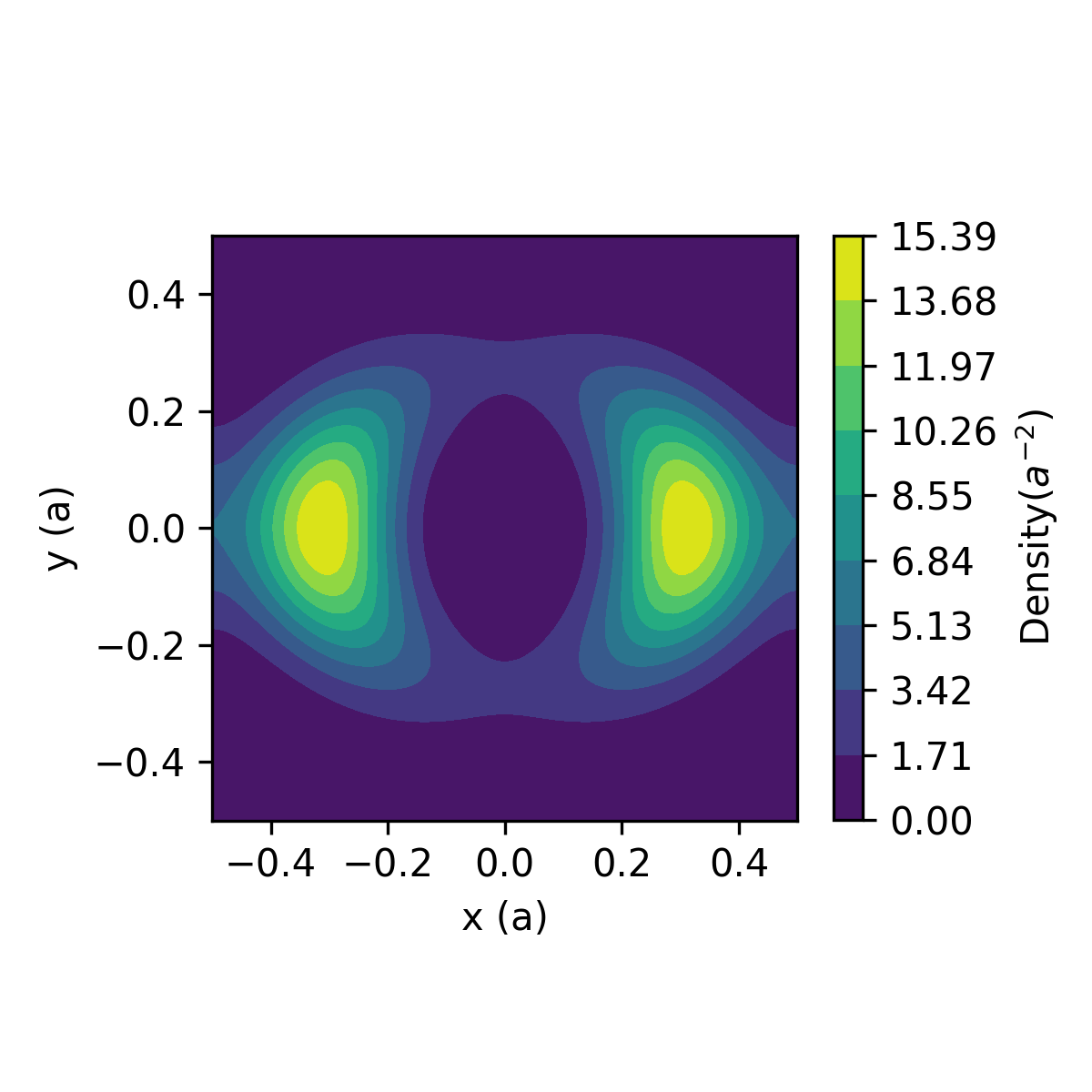}
    \end{subfigure}
    \\
    \begin{subfigure}
        \centering
        \includegraphics[trim=10 40 25 40, clip, width=0.31\textwidth]{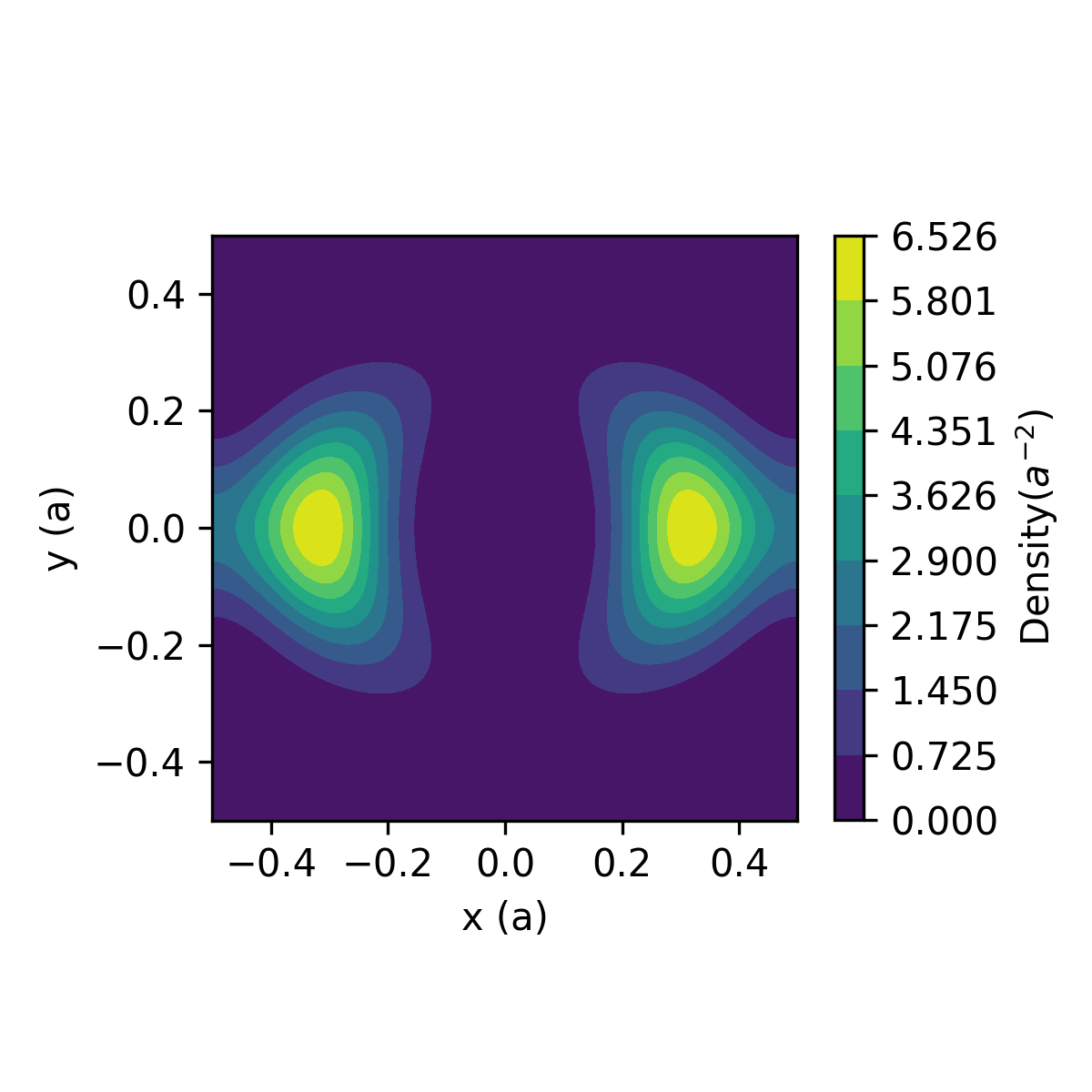}
    \end{subfigure}
    \quad
    \begin{subfigure}
        \centering
        \includegraphics[trim=10 40 25 40, clip, width=0.31\textwidth]{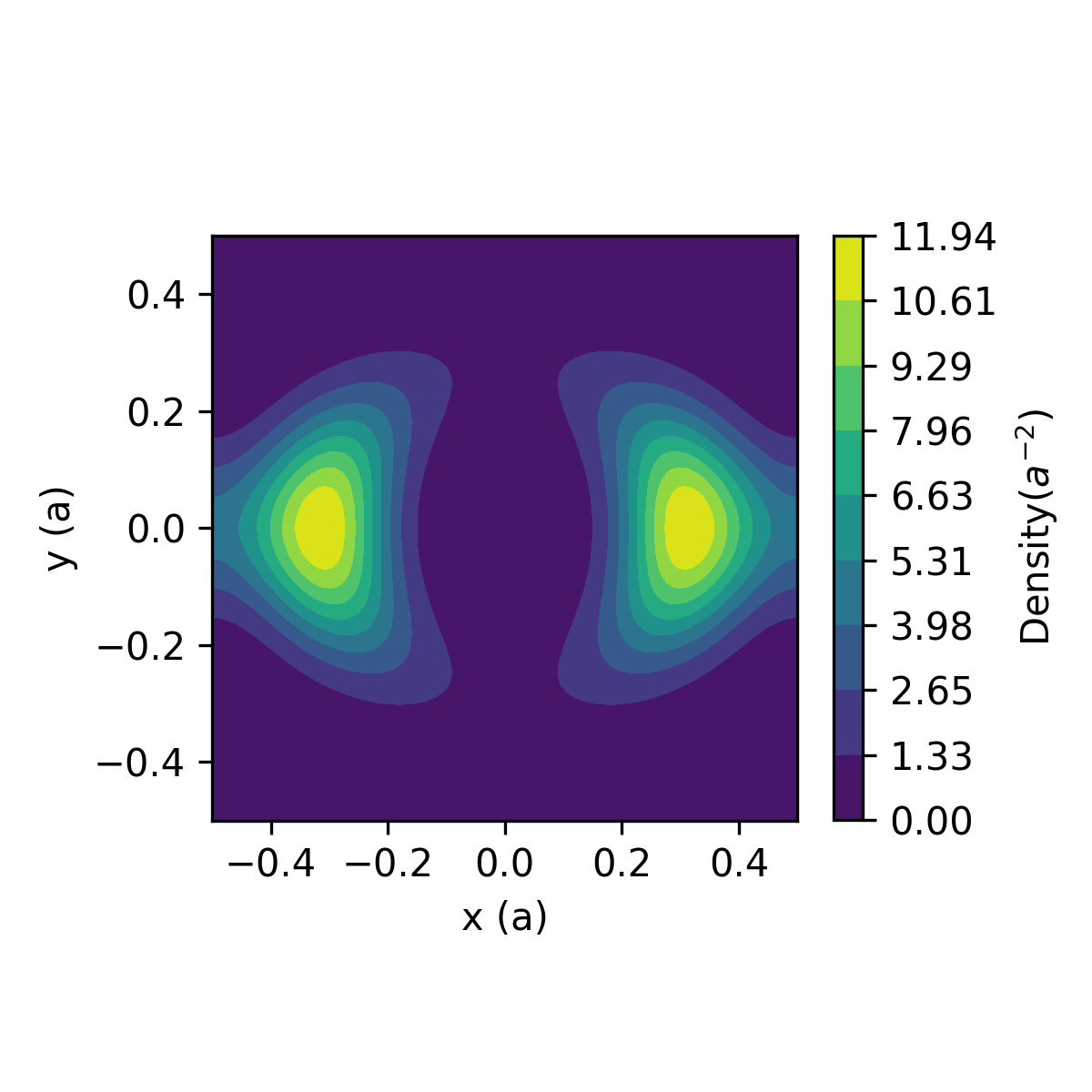}
    \end{subfigure}
    \quad
    \begin{subfigure}
        \centering
        \includegraphics[trim=10 40 25 40, clip, width=0.31\textwidth]{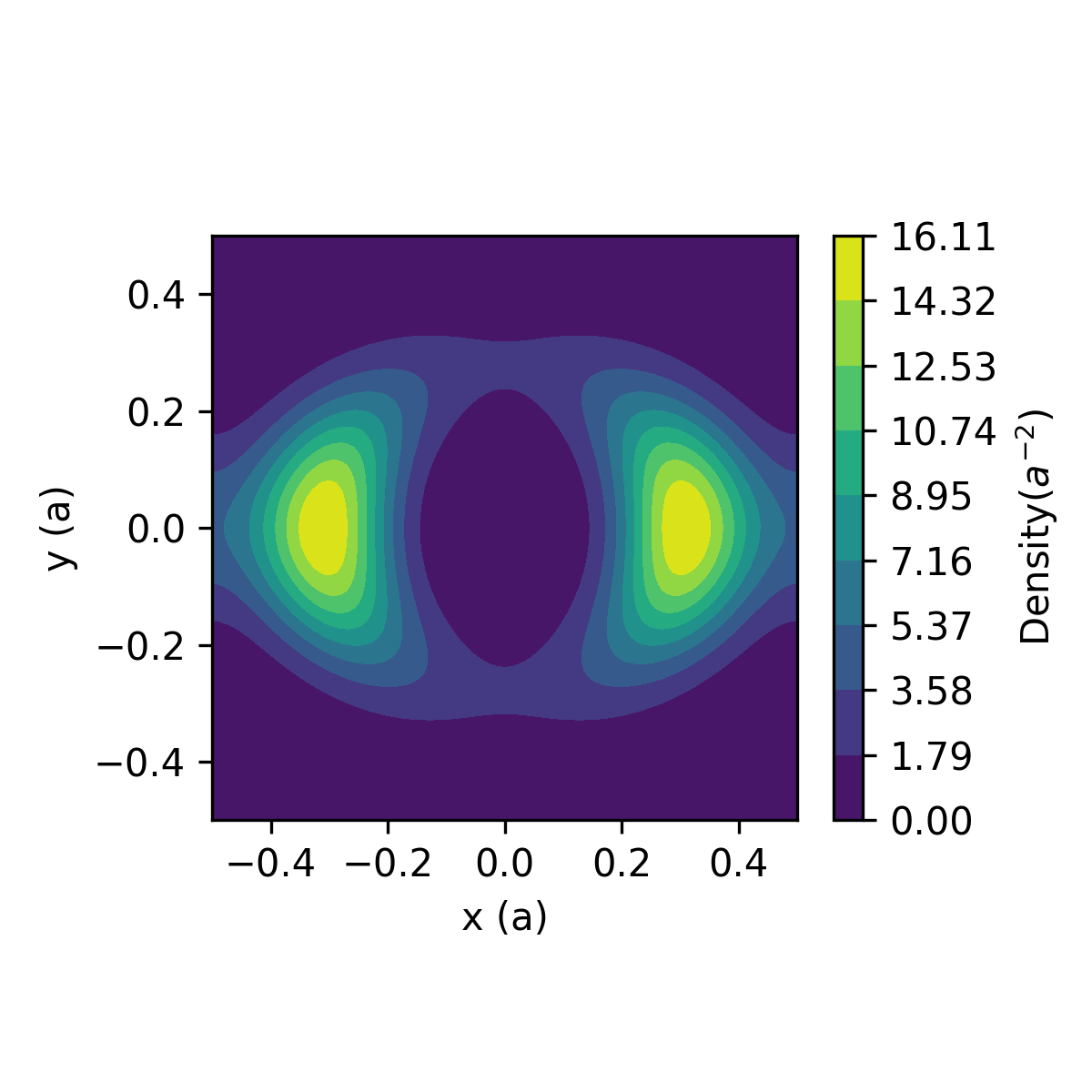}
    \end{subfigure}
    \\
    \begin{subfigure}
        \centering
        \includegraphics[trim=10 40 25 40, clip, width=0.31\textwidth]{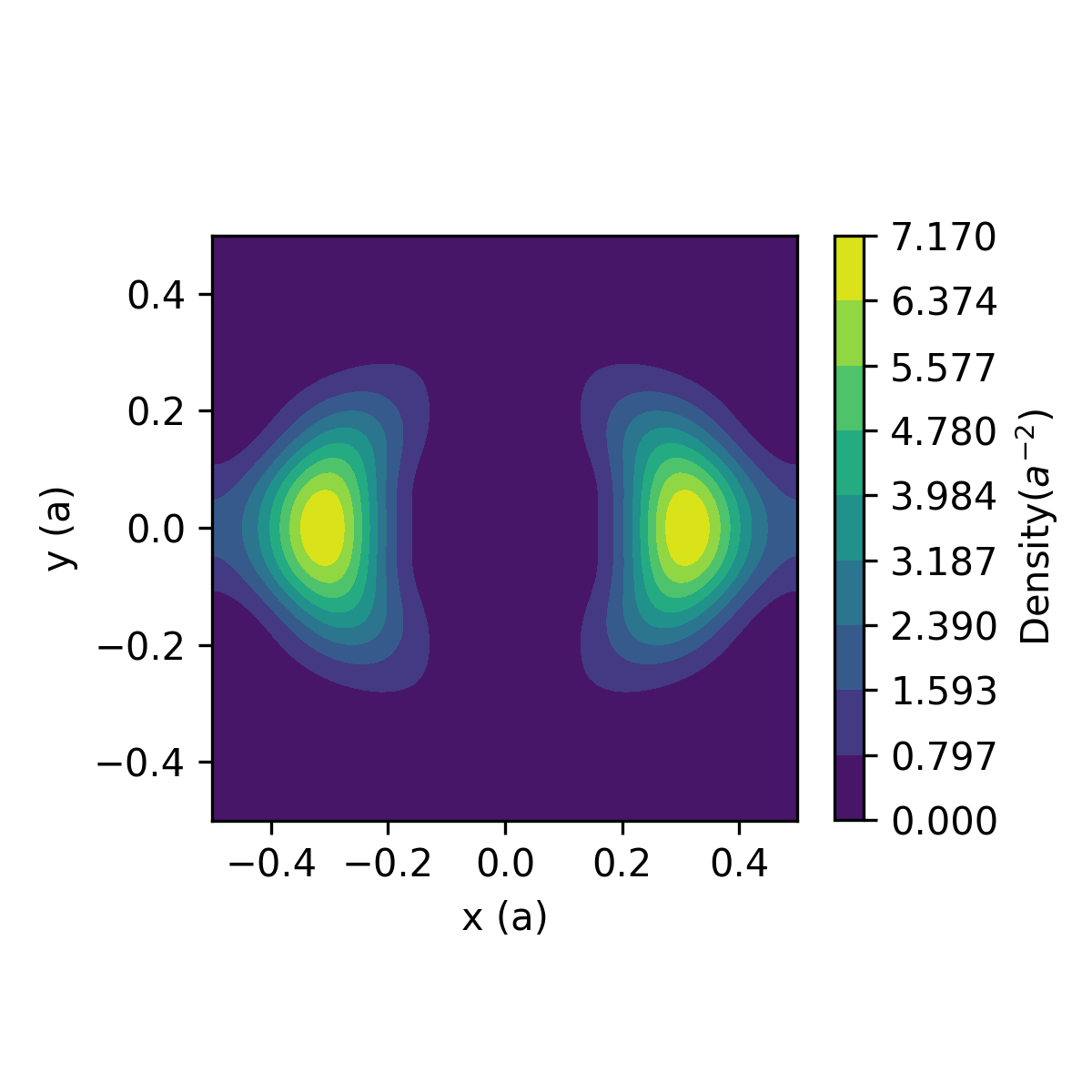}
    \end{subfigure}
    \quad
    \begin{subfigure}
        \centering
        \includegraphics[trim=10 40 25 40, clip, width=0.31\textwidth]{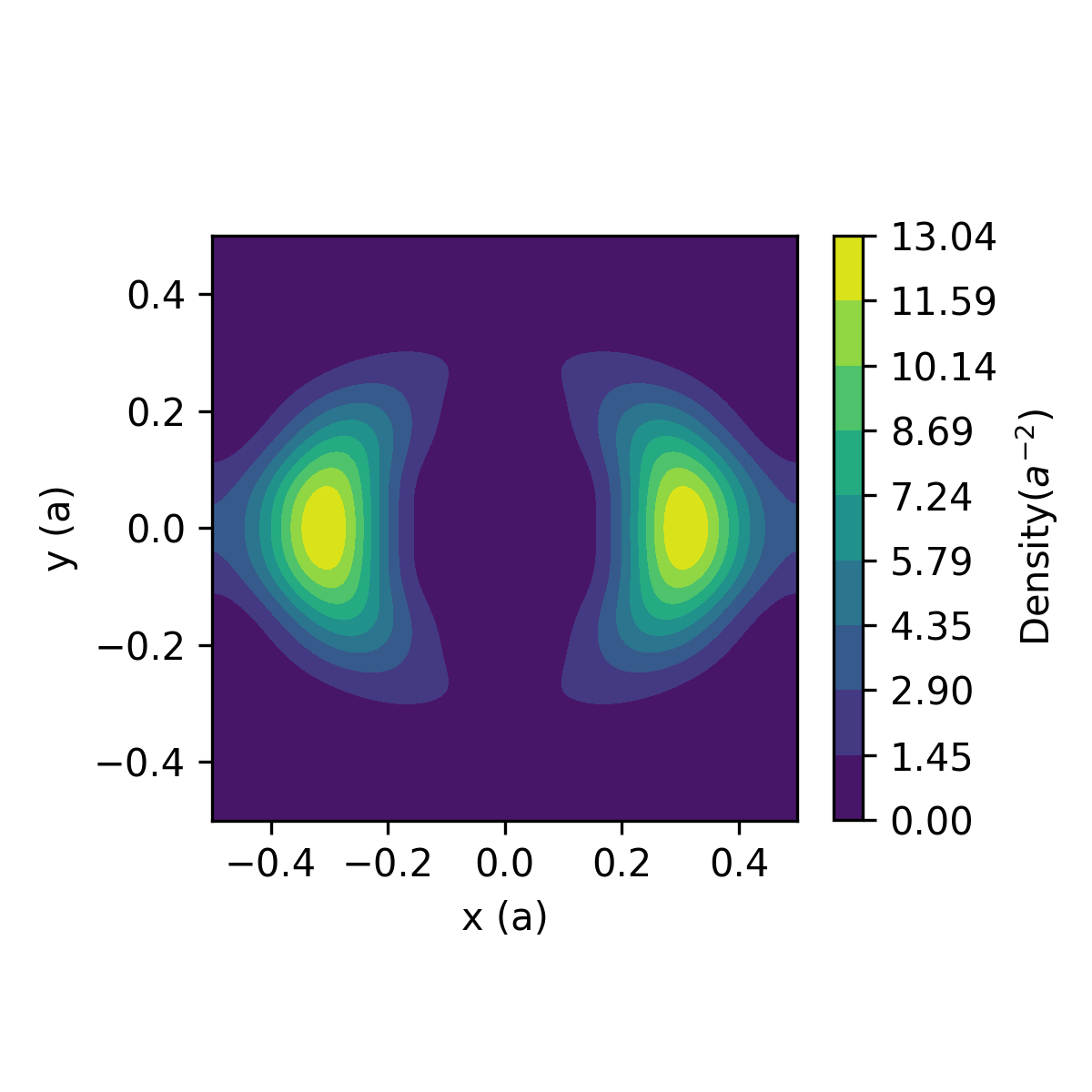}
    \end{subfigure}
    \quad
    \begin{subfigure}
        \centering
        \includegraphics[trim=10 40 25 40, clip, width=0.31\textwidth]{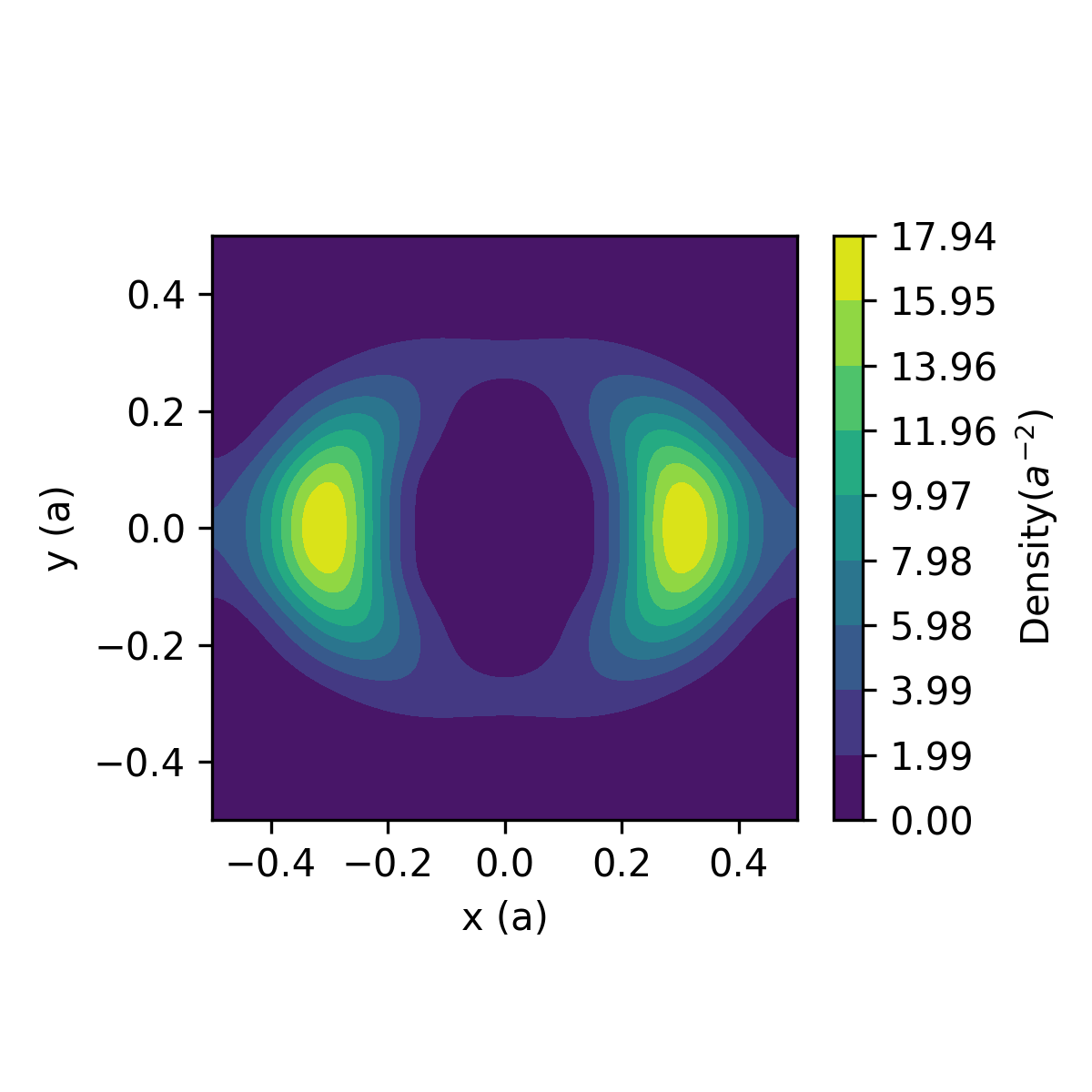}
    \end{subfigure}
    \caption{Electron density distribution for one (left column), two (the middle column) and three (the right column) electrons per unit cell. The rows from up to down correspond to $B=10$T, $B=13.65$T and $B=22.55$T, respectively.}
    \label{dens}
\end{figure*}

\begin{figure*}
    \centering
    \begin{subfigure}
        \centering
        \includegraphics[trim=5 20 40 30, clip, width=0.31\textwidth]{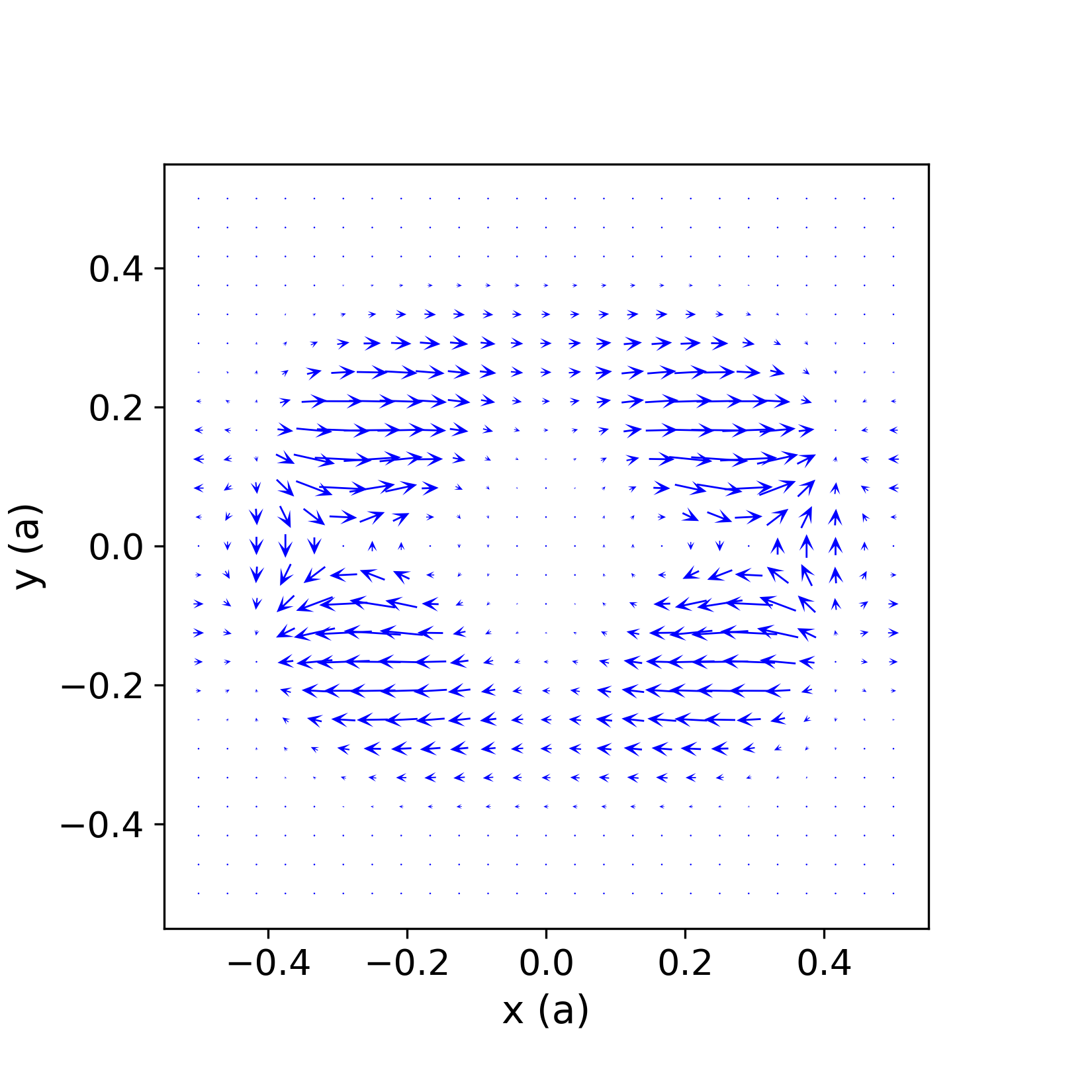}
    \end{subfigure}
    \quad 
    \begin{subfigure}
        \centering
        \includegraphics[trim=5 20 40 30, clip, width=0.31\textwidth]{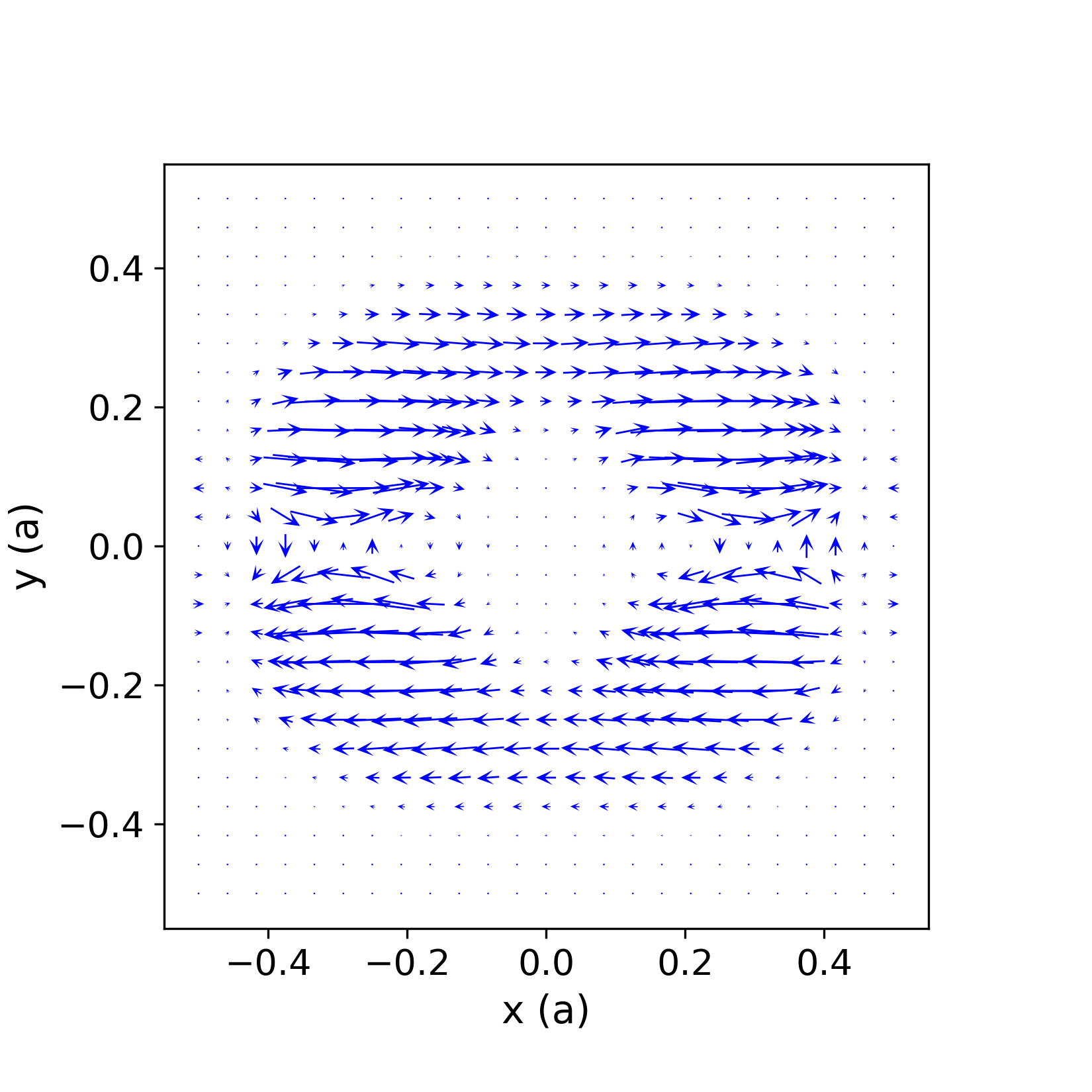}
    \end{subfigure}
    \quad
    \begin{subfigure}
        \centering
        \includegraphics[trim=5 20 40 30, clip, width=0.31\textwidth]{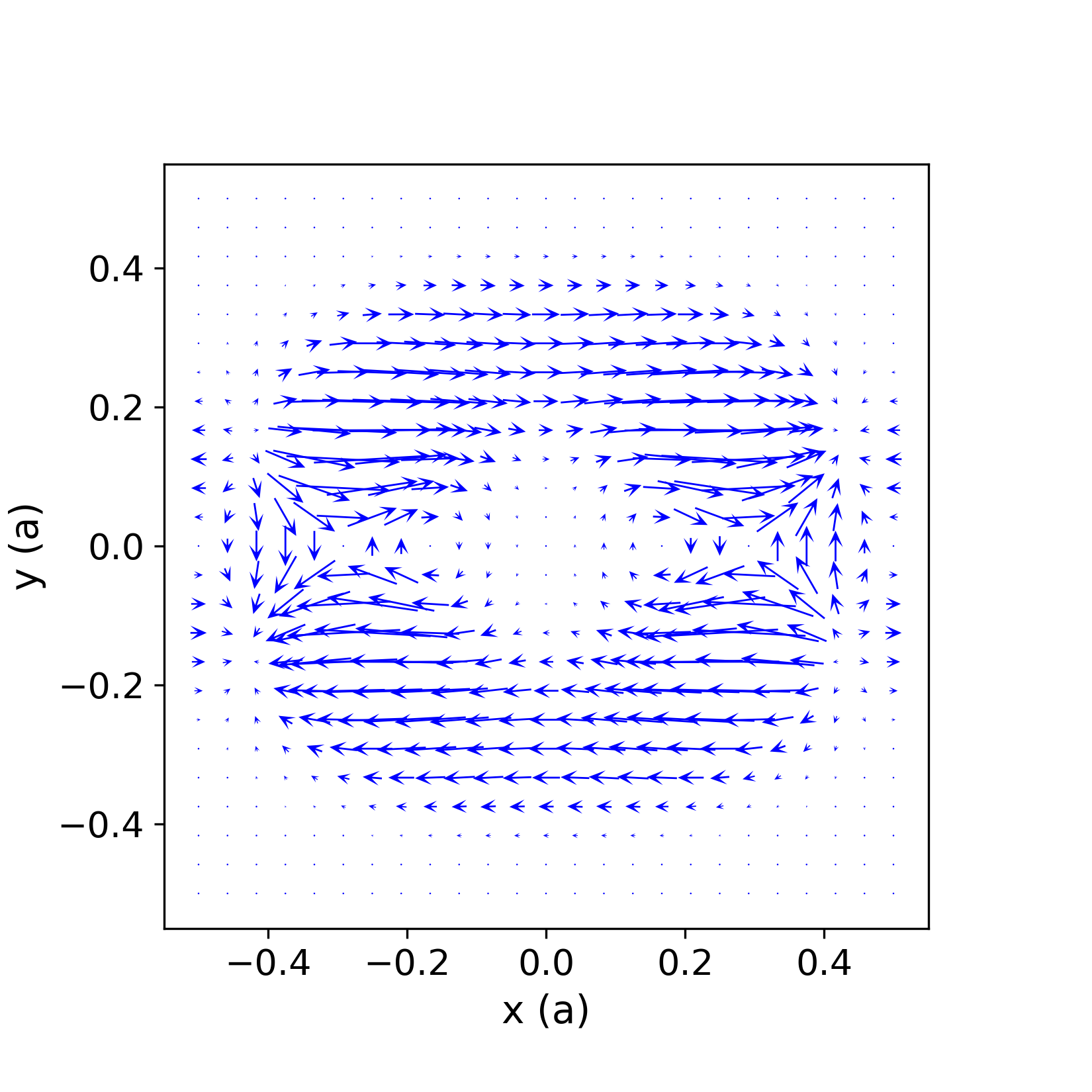}
    \end{subfigure}
    \\
    \begin{subfigure}
        \centering
        \includegraphics[trim=5 20 40 30, clip, width=0.31\textwidth]{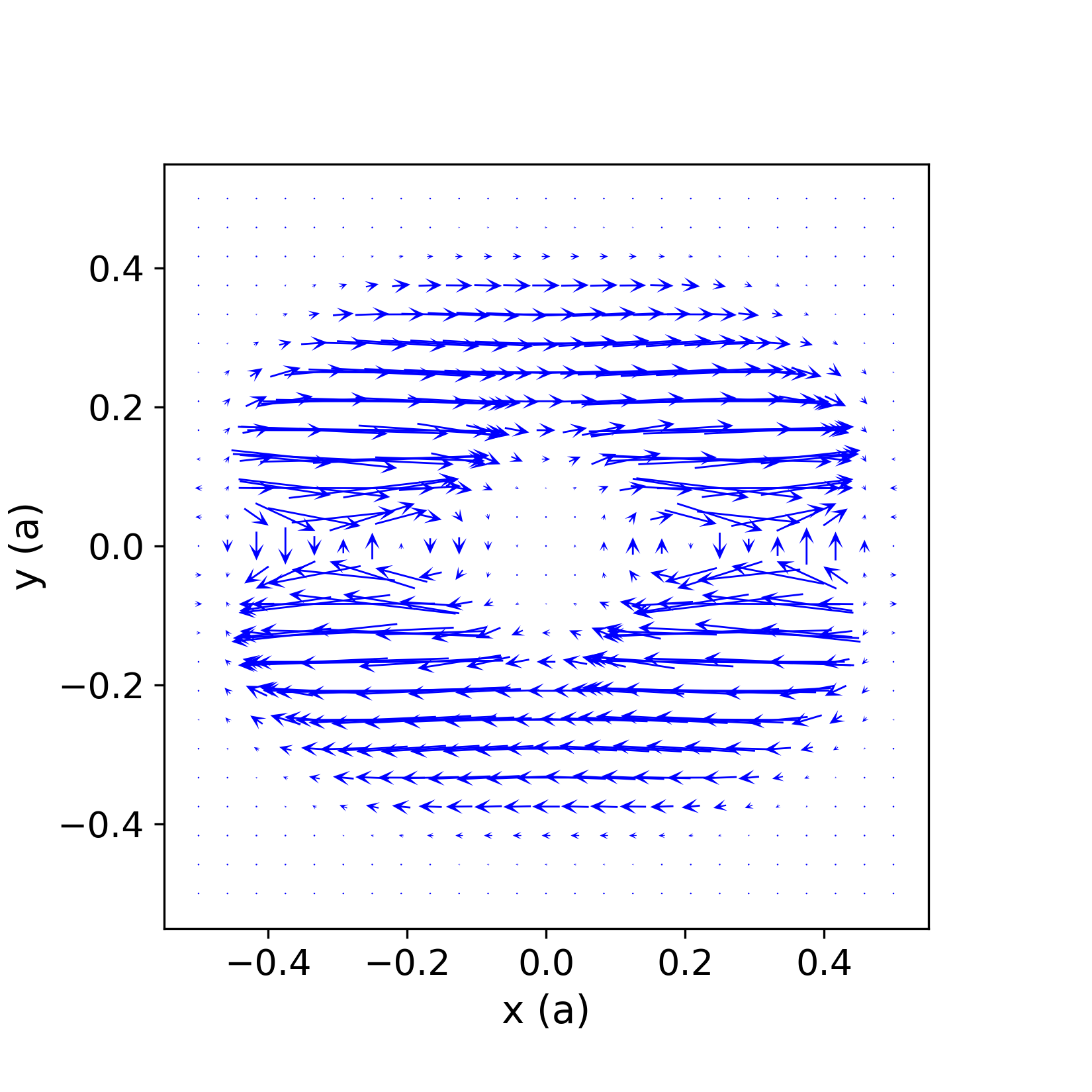}
    \end{subfigure}
    \quad
    \begin{subfigure}
        \centering
        \includegraphics[trim=5 20 40 30, clip, width=0.31\textwidth]{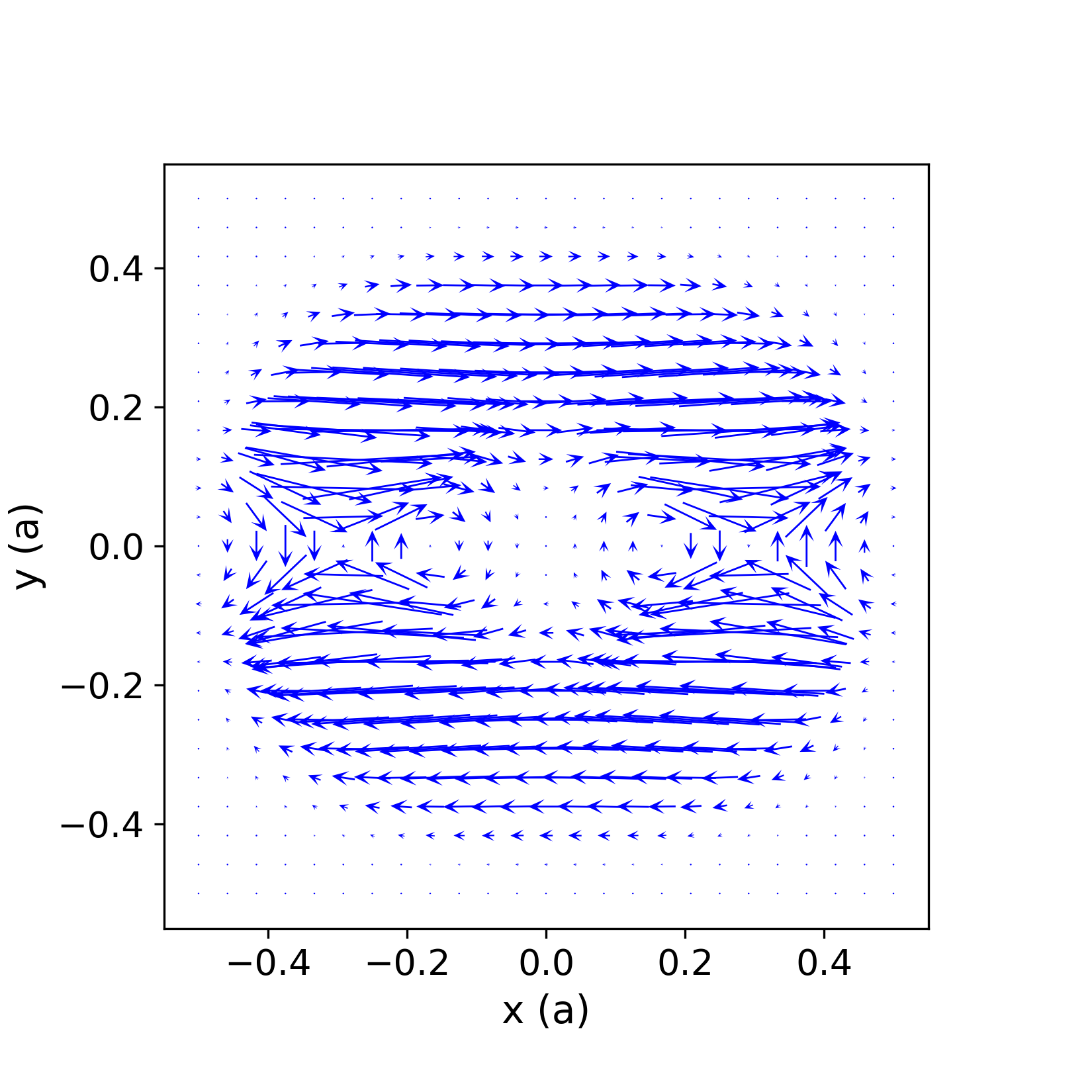}
    \end{subfigure}
    \quad
    \begin{subfigure}
        \centering
        \includegraphics[trim=5 20 40 30, clip, width=0.31\textwidth]{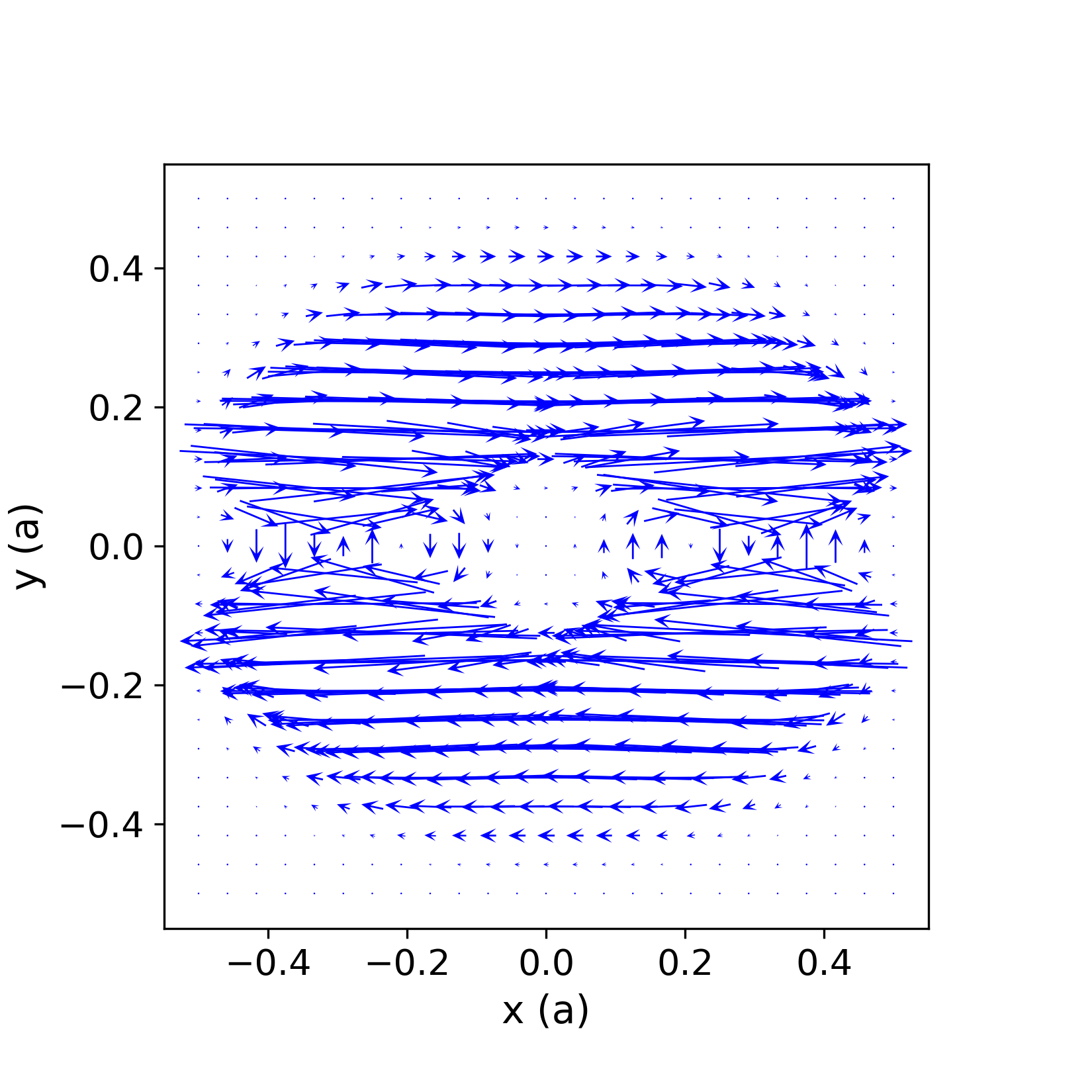}
    \end{subfigure}

    \caption{Persistent current density distribution for one (left column), two (the middle column) and three (the right column) electrons per unit cell of the chain. The rows from up to down correspond to $B=13.65$T and $B=22.55$T, respectively.}
    \label{current}
\end{figure*}
The numerical calculations are carried out for GaAs material for the following values of parameters: $a = 42$ nm, $V_0 = 250$ meV, $m=0.067m_0$ ($m_0$ is the free electron mass), $g^{*}=-0.44$, and $\varkappa = 12.4$.
We consider a low but non-zero temperature $T = 4$ K, which corresponds to thermal energy of about $0.342$ meV.
Figs.\ \ref{Enonint1} and \ref{Enonint2} represent the dependencies of the non-interacting electron energies on the magnetic field for the lowest two and the next four minibands, respectively.
The solid lines correspond to spin-up states while the dashed ones are for spin-down states.
The overall behavior of the curves, as well as the existence of miniband nodes for certain values of magnetic field are in agreement with the previously obtained results for a chain of QRs with a rectangular potential profile \cite{MANSOURY2022128324}.
The qualitative coincidence of the results is caused by the fact that the origin of the miniband nodes is connected with the geometry of the structure and the topology of QRs rather than the shape of the modulation profile.
The small contribution of the Zeeman energy is connected to the small value of the Land\'{e} factor for GaAs.
The curves inside each miniband correspond to eleven different values of quasi-momentum, namely $ka = \pm (\pi/5) n$ for integer values of $0 \leq n \leq 5$.
The appearance of six different values of energy in each miniband is due to the doubly degenerate states with regard to the direction of the quasi-momentum.
In contrast to our previous result, we observe here two detached minibands with nodes almost for the same values of $B$ (Fig.\ \ref{Enonint1}).
These minibands are composed of bound-like states as they correspond to energy values which are below the values of potential (around $0$, see Eq.\ (\ref{Vpot})) between the QRs.
The origin of the nodes can be elucidated within the framework of the following general comprehension.
When moving from a unit cell of the chain to the next one the electron wave function gains a phase $\Delta \phi = ka + (e/\hbar c)\int \vec{A}d\vec{l}$, where the integration path depends on quasi-momentum $k$.
In the absence of the modulation potential the mean value of $y$ coordinate $\langle y \rangle = l_{B}^{2} k$ (see the formulas (7) and (8)) meaning that $ \Delta \phi = 0$.
In this case no dispersion is observed and one deals with discrete Landau levels.
In the presence of the modulation potential a vanishing dispersion means that $\Delta \phi = 2 \pi n$ with integer $n$, or $ka - \langle y \rangle a / l_{B}^{2} = 2 \pi n$, where $\langle y \rangle$ is the mean value of the $y$ coordinate for a given state.
As $\langle y \rangle / l_{B}^{2}$ is defined by the value of magnetic field and the number of the miniband, the later condition leads to repeating nodes with increasing $B$, which are not exactly periodic and which are differently positioned in different minibands.
It is clear, that the 3rd and the 4th minibands undergo crossings, while there are obvious anticrossings between the 4th and the 5th minibands (Fig.\ \ref{Enonint2}).
One can also clearly note, that the minibands are wider in the regions of energy decrease (with increasing $B$), but they are extremely narrow in the regions of energy increase with the nodes emerging in the latter regions.

Fig.\ \ref{Eint} represents the energy dependencies for the first (the blue lines) and the second (the orange lines) spin-up minibands on the magnetic field for the Hartree-interacting electrons.
On the left (right) panel of the figure the curves from down to up correspond to $1$, $2$ and $3$ ($0.5$ and $1.5$) electrons per unite cell (UC), respectively.
We have plotted the case of non-interacting electrons (light-blue lines) as well for comparison. The vertical lines indicate the positions of nodes for non-interacting electrons in the first (blue vertical lines) and the second (red vertical lines) minibands.
The values $B=13.65$T and $B=22.55$T correspond to the nodes of the 1st miniband, while $B=14.375$T and $B=22.9$T correspond to ones for the 2nd miniband.
For interacting electrons we have not considered the the values of $B < 10$ T as it would lead to a significant increase in the Hamiltonian matrix size and to a time-expensive Hartree-iterations. But in principle, one can do the calculations for any values of $B$.
It is clear that minibands shift up with increasing number of electrons per UC $N_e$ due to their repulsion and due to the smoothening of the effective potential $V_{\mathrm{ext}}(\vec{r})+V_\mathrm{H}(\vec{r})$.
Higher the miniband, broader it is away from the nodes.
A more detailed observation indicates a slight shift of the nodes' positions to the higher values of $B$ due to the increase in electrons' number per UC (see the insets of the figures).
This is connected with a larger value of $\langle y \rangle$ for an electron traveling in between UCs due to a weaker localization in the $y$ direction.

The electron density distribution in a UC of the QR chain for 1 (the left column), 2 (the middle column) and 3 (the right column) electrons per UC is presented in Fig.\ \ref{dens}.
The rows from top to bottom correspond to $B=10$T, $B=13.65$T and $B=22.55$T, respectively.
In all the cases the electrons are mainly located around the axis $x$ because of the strong confinement in the $y$ direction.
The increase of the magnetic field leads to even stronger localization, while the increase of the electrons' number per a UC results in a more ring-like distribution of the electrons, which is a direct consequence of the screening.

Fig.\ \ref{current} demonstrates the vector plot for the persistent current density in a UC of the QR chain.
As in the previous figure the left, middle and the right columns correspond to the number of electrons $N_{e}=1$, $N_{e}=2$, and $N_{e}=3$, respectively.
The upper row is plotted for $B=13.65$T and the lower row is for $B = 22.55$T.
It is clear that both the increase in the magnetic field and in the number of electrons result in stronger currents in the QR region.
Furthermore, both of the aforementioned factors contribute to a more pronounced curvilinear motion of the electrons.
However the total current is zero, as is expected for a system in magnetic field in equilibrium.
One can see that in the upper half of the system with $y \geq 0$ contributes mainly in the current directed to the right, while the opposite happens for $y \leq 0$.
There is no current in the points where the electronic density has a maxima or a minima, in other words, where the density gradient is $0$, which is in accordance with the general understandings.
In addition there are some points around which one can see clear circulating currents (for example the points with $x \approx \pm 0.2$ and $y=0$).

\begin{figure}
    \includegraphics[trim=0 40 10 40, clip, width=0.37\textwidth]{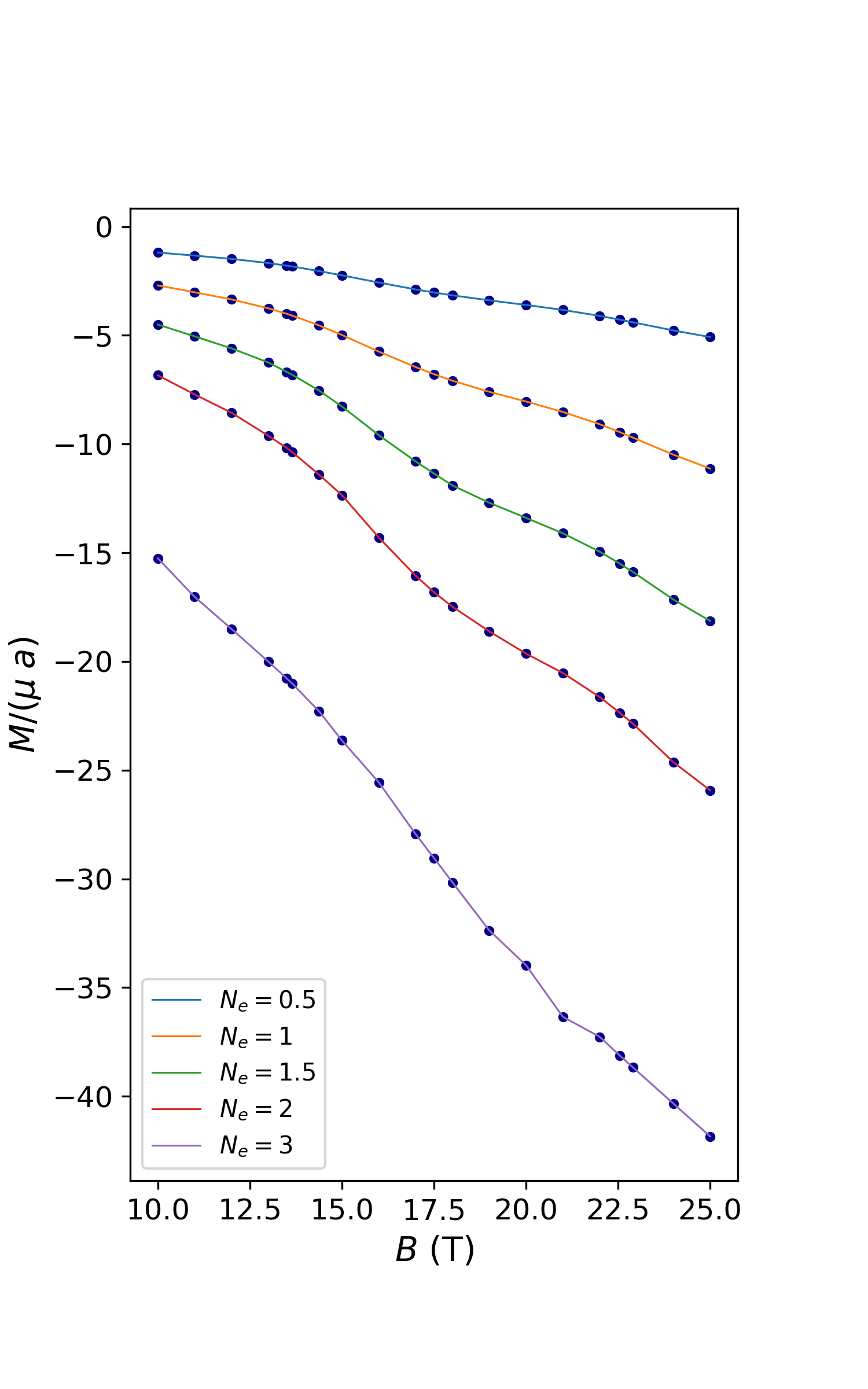}
    \caption{Dependence of the QR chain magnetization (for a UC in the unites of the effective Bohr magneton) on the magnetic field induction for different numbers of electrons per a UC.}
    \label{magnetization}
\end{figure}

The magnetization of the system versus magnetic field is shown in Fig.\ \ref{magnetization}.
An obvious diamagnetic response is observed, which significantly depends on the number of electrons per a UC.
It can be seen from the figure that the curves have slight oscillations connected to the sequential appearance of the miniband nodes (Fig. \ref {Eint}), where a
stronger contribution of rotational currents in the magnetization takes place.

\begin{figure}
        \includegraphics[trim=0 40 10 40, clip, width=0.37\textwidth]{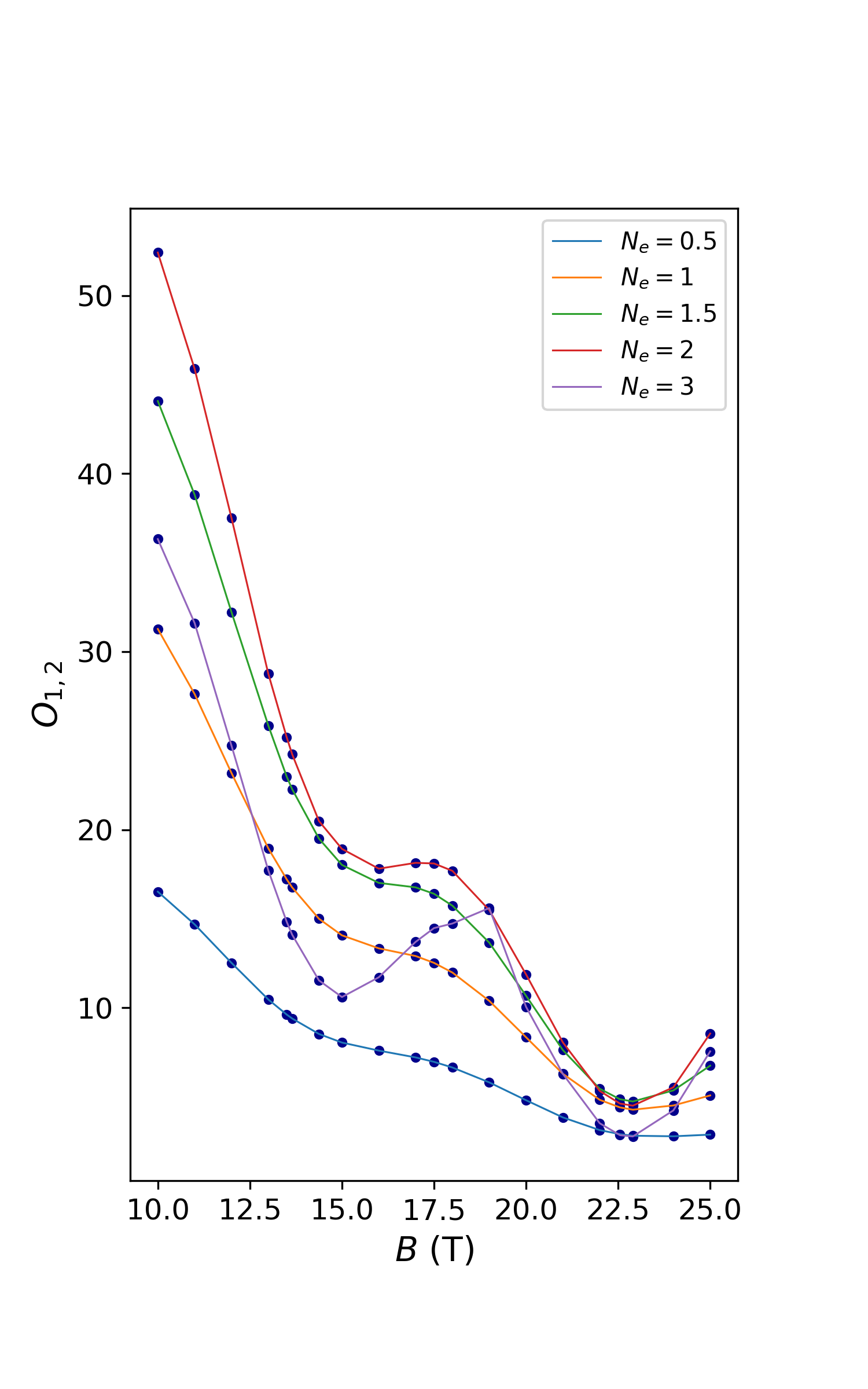}
    \caption{Dependence of the oscillator strength between the first and the second minibands of the QR chain on the magnetic field induction for different numbers of electrons per a UC.}
    \label{oscillator}
\end{figure}

Fig.\ \ref{oscillator} represents the dependencies of the oscillator strength between the two lowest Zeeman-couples of minibands on the magnetic field for different numbers of electrons per a UC.
One can notice obvious local minima of the oscillator strength around the values of $B$ corresponding to the miniband nodes.
The overall decrease of the oscillator strength is due to the restriction of the energy difference between the minibands with increasing $B$.
The curves shift up with the increase of the number of electrons per UC while $N_{e} \leq 2$.
The further increase in $N_{e}$ leads to the shift of the curves down.
The reason is that for $N_{e} \geq 2$ the second Zeeman-couple of minibands starts to be filled resulting to a smaller transition probability.
One can also note more significant oscillations of the curve corresponding to $N_{e} = 3$ compared to other cases considered here.
We consider non-zero, yet very small value for the temperature.
Even a very small thermal energy is enough to populate partially all the states inside a miniband where the chemical potential lies if we consider magnetic field values near a miniband node.
On the other hand, the partially filled final states have smaller contribution to the oscillator strength comparing with empty ones.

\section{Summary}
\label{summary}
In summary, we have considered the equilibrium properties and interminiband transitions for Hartree-interacting 2D EG in a 1D chain of planar QRs subjected to a transverse homogeneous magnetic field.
We provide an analytical model for the external potential which reflects three important properties of the considered structure: the periodicity in the $x$ direction, the confinement in the $y$ direction and a ring topology in each UC of the chain.
Our calculations of the electronic band structure versus magnetic field show the existence of two detached minibands in the region of smaller energies, while higher minibands manifest multiple crossings and anticrossings which are analogues of the Aharonov-Bohm crossings for a single QR.
The combined effect of the translational and magnetic phases gained by the electron when traveling between neighboring UCs results in highly degenerate energy levels (referred to as miniband nodes) for certain quasi-periodic values of magnetic field.
The miniband nodes are preserved when taking into account the Hartree-type Coulomb interaction between electrons.
However, the interaction leads to an up-shift and a broadening of the minibands, as well as to a slight shift of the nodes toward larger values of magnetic field.
We have shown that miniband nodes have their signature in the magnetization of the EG and interminiband transitions.
The dependence of the magnetization on the magnetic field reveals slight oscillations around the nodes, while the oscillator strength displays significant minima.
The number of electrons per UC has a strong impact on the current density distribution, magnetization and the oscillator strength.
The magnetization decreases, while the oscillator strength has a non-monotonic dependence on the electrons' number.
The obtained results are of significant interest in view of flexible manipulation of magneto-optical properties of future devices operating in the far-infrared and THz regimes.

\section{Acknowledgements}
\label{acknowledgements}
This work was financially supported by the Armenian State Committee of Science (grants No 21SCG-1C012, 24WS-1C040 and 21AG-1C048),
by the Research Fund of the University of Iceland grant No. 92199, and the Icelandic Infrastructure Fund. The computations were performed in the Center of Modelling and Simulations of Nanostructures at Yerevan State University.
\end{document}